# Operation regimes, gain dynamics and highly stable operation points of Ho:YLF regenerative amplifiers

PETER KROETZ,[1,2,4,*] AXEL RUEHL,[3] ANNE-LAURE CALENDRON,[1,4,5] GOURAB CHATTERJEE,[1,2] HUSEYIN CANKAYA,[1,3,5] KRISHNA MURARI,[1,3,4] FRANZ X. KÄRTNER,[1,3,4,5] INGMAR HARTL,[3] AND R. J. DWAYNE MILLER[1,2,5,6]

[1]*Center for Free-Electron Laser Science (CFEL), Notkestraße 85, D-22607 Hamburg, Germany*
[2]*Max-Planck Institute for the Structure and Dynamics of Matter (MPSD), Luruper Chaussee 149, 22761 Hamburg, Germany*
[3]*Deutsches Elektronen-Synchrotron (DESY), Notkestrasse 85, 22607 Hamburg, Germany*
[4]*Department of Physics, University of Hamburg, 22761 Hamburg, Germany*
[5]*Centre for Ultrafast Imaging (CUI), Universität Hamburg, Luruper Chaussee 149, 22761 Hamburg, Germany*
[6]*Departments of Chemistry and Physics, University of Toronto, Toronto M5S 1A7, Canada*
*\*peter.kroetz@mpsd.mpg.de*

**Abstract:** We present a comprehensive study of laser pulse amplification with respect to operation regimes, gain dynamics, and highly stable operation points of Ho:YLF regenerative amplifiers (RAs). The findings are expected to be more generic than for this specific case. Operation regimes are distinguished with respect to pulse energy and the appearance of pulse instability, and are studied as a function of the repetition rate, seed energy and pump intensity. The corresponding gain dynamics are presented, identifying highly stable operation points related to high gain build-up during pumping and high gain depletion during pulse amplification. These operation points are studied numerically and experimentally as a function of several parameters, thereby achieving, for our Ho:YLF RA, highly stable output pulses with measured fluctuations of only 0.19 % (standard deviation).

## 1. Introduction

Pulse amplification with regenerative amplifiers (RAs) is one of the most common approaches to amplify low energetic laser pulses to multi-mJ pulse energies. Important for the design of any RA is its ability to generate energetically stable output pulses. Initially, observed instabilities in the output pulses from an RA were attributed to inherent noise originating from the pump sources [1]. However, later it was reported that exceeding the repetition rate beyond a certain threshold value can also be a potential cause for large energy fluctuations [2,3]. This phenomenon, called bifurcation (or multifurcation), can cause output pulses to fluctuate between two (or more) values, and is an effect of the gain dynamics in RAs for consecutive pump and amplification cycles [3].

Although energetically stable output pulses are generally important for many applications, the basic understanding of the onset of bifurcation based on RA parameters such as pump intensity, seed fluence and the number of round trips (RT), as well as the underlying gain dynamics remains incomplete. Similarly, even for RA operation at a single-pulsing state (meaning in absence of large scale pulse instability), there are no comprehensive studies and guidelines for an optimized noise performance, to the best of our knowledge.

The first experimental and numerical work dealing with the onset of bifurcation and chaotic output pulses was published by Dörring et al. [3] for a Yb:glass RA. The susceptibility of RAs to show unstable output pulses was connected to the repetition rate of the RA and the upper state lifetime of the gain medium ($\tau_{gain}$). For repetition rates on the order of the inverse lifetime $1/\tau_{gain}$, RAs can show unstable output pulses. It was later shown that the seed energy also affects the onset of bifurcation. Higher seed energies allow operation at higher repetition rates without the presence of bifurcation [4]. The first experimental hint that there is a single-pulsing state beyond the bifurcation instability, corresponding to higher RTs for a fixed pump power, was demonstrated in [3]. This was more precisely demonstrated for an Nd:YVO$_4$ RA, showing bistable output pulses for a large range of RTs, prior to the start of a further single-pulsing state [5]. More recently, this was demonstrated for the first time for an Ho:YLF RA, achieving high pulse energies by overcoming bifurcation instability at a repetition rate of 1 kHz [6]. Furthermore, our measurements and simulations showed that the output pulse stability is highest at a specific RT close to the maximum average output pulse energy, independent of the presence of bifurcation in the system at an earlier RT [6].

RAs lasing at a wavelength of 2 μm and based for example on Ho:YLF, Ho:YAG [7] and Tm:YAP [8] have recently garnered attention as a viable driving source for mid-IR optical parametric amplifiers (OPA) [9]. They can be used in conjunction with highly efficient non-oxide nonlinear crystals (such as ZnGeP$_2$ [9,10]), as opposed to the more conventional 1-μm pump sources. Unfortunately, Ho:YLF RAs, with an upper state lifetime of about 15 ms (or equivalently, an inverse lifetime of ∼ 67 Hz), greatly suffer from the onset of bifurcation at the demonstrated repetition rates. Typically, Ho:YLF RAs were either operated at relatively low repetition rates [9], thus completely suppressing the onset of bifurcation, at high repetition rates until the onset of bifurcation [11], or directly in the bifurcation at a stable double-pulsing state [12]. In the latter case, pulse-picking only the higher energy pulse at half the repetition rate, allows the extraction of stable and high-energy pulses [12,13].

By employing high pump intensities, we demonstrated operation of our Ho:YLF RA up to repetition rates of 750 Hz without any sign of bifurcation, which is more than an order of magnitude higher than the inverse lifetime [6]. This indicates that the pump intensity also plays a crucial role in the onset of bifurcation. However, the previously used numerical model and related study of the RA gain dynamics consider the gain build-up during the pumping phase rather qualitatively as a function of the lifetime and neglect an explicit dependence of the pump intensity on the gain build-up [3,4].

In this paper, we show a comprehensive study of operation regimes, gain dynamics, and highly stable operation points of CW-pumped Ho:YLF RAs. We numerically distinguish four different RA operation regimes exhibiting different output pulse characteristics as a function of the pump intensity, repetition rate and seed fluence. These operation regimes represent inherent limitations in RA amplification in terms of stable output pulse energy and average output power, that are crucial for any RA user and developer. We analyze the onset of bifurcation and we empirically find that for high gain RAs the repetition rate at which bifurcation can appear scales linearly with the pump intensity and follows a power-law with the seed energy.

We furthermore present the gain dynamics for the RA operation regimes in terms of the normalized gain and gain depletion. We identify an operation point that offers the highest output pulse stability, located at high normalized gain and gain depletion values. We analyze such highly stable operation points numerically and experimentally as a function of various RA parameters. We thus achieved with our laboratory Ho:YLF RA highly stable output pulses with measured energy fluctuations of only 0.19 % (standard deviation). Although this study is conducted with an Ho:YLF RA, the results are considered more generic and provide a more complete and general understanding of pulse amplification in RAs, independent of the gain material.

The paper has been structured in the following manner: Section 2 describes the simulation model that is used for this study. In section 3, the different RA operation regimes are distinguished and discussed, and the onset of bifurcation is studied as a function of the pump intensity and seed fluence. Section 4 describes the gain dynamics for the operation regimes and identifies a highly stable operation point. Analogous highly stable operation points are studied numerically and experimentally as a function of various RA parameters in section 5.

## 2. Numerical simulation method

This chapter presents the general simulation procedure used for the simulations in this paper. The model simulates consecutive pumping and pulse amplification cycles (sketched out in Fig. 1), which allows a statistical analysis of a large ensemble of amplified output pulses. At the beginning of the simulations, a random value for the initial inverted fraction $\beta$ in the laser crystal can be used as a starting value. During the pumping and amplification cycles, the inverted fraction $\beta$ increases or decreases and the final inverted fraction $\beta^p$ or $\beta^a$ is fed as input for the following amplification or pumping cycle, respectively. Noise originating from the pump and seed source is included by varying the pump and seed fluences for consecutive
pumping and amplification cycles according to a Gaussian distribution. In this paper, we refer to the noise of the pump and seed source as the standard deviation of consecutive pump and seed fluences in percent of the corresponding mean fluence. Equivalently, the standard deviation of the output pulse fluence in percent of the mean output fluence is used as a measure for the output noise.

To decouple the simulation results from the starting conditions, we neglect a certain number of pumping and amplification cycles in the analysis. This number depends on the type of analysis conducted. The simulations to distinguish RA operation regimes were performed without pump and seed noise and show the inherent susceptibility of the RA to show instable output pulses. We observe that simulating 10 pump and amplification cycles and neglecting the first 5 cycles generally allows distinguishing the operation regimes. To statistically study the noise of the output pulses when pump noise is included, a larger number of pumping and amplification cycles is simulated. Here we always simulate 1500 cycles and neglect the first 100 cycles.

The simulation framework was already presented in detail in [14], where the model was spectrally generalized and used to study spectral shaping effects in Ho:YLF RAs. Here, we used the monochromatic version as it is computationally less expensive. Although the simulation model is quite simplistic, it captures the essential physics necessary for this study. The model does not consider for example temporal or spatial effects, nor amplified spontaneous emission (ASE), up-conversion, thermal and non-linear effects. Furthermore, although CW-pumping is assumed, the model neglects a pumping effect during the typically very short pulse-amplification cycle. However, despite these simplifications, this model was already successfully employed to reproduce the results from our Ho:YLF RA, including the onset and end of bifurcation instability, and the existence of a highly stable operation point close to gain depletion [6]. The simulation model utilizes the classical monochromatic Frantz-Nodvik formalism (FN) [15], which is a common approach to simulate pulse amplification in a laser gain media, as long as the amplification process is short compared to the upper state lifetime of the excited ions. The classical FN equations are comprehensively explained in [16-18]. To simulate the pumping process with this set of equations, we additionally correct the inverted fraction with the inversion decay law [19] during the pumping time.

Equations (1)-(5) describe a single-pass of an incoming fluence through the gain medium. The fluence can be either amplified or absorbed (depending on the emission and absorption cross sections $\sigma_{em}$ and $\sigma_{abs}$ at the considered wavelength) and the inverted fraction decreases or increases from $\beta_{i-1}$ to $\beta_i$, respectively, as calculated by Eq. (6)-(9). The correction term to account for the inversion decay [Eq. (9)] typically can be neglected for amplification and pumping processes that are much shorter than the upper state lifetime.

The incoming pump fluence is temporally divided by slicing it into smaller fluence slices with a time duration of $\Delta t$ each. Each slice passes through the gain medium separately, and the inverted fraction is consecutively updated. More details about the updating scheme can be found in [14].

The single pass gain of a quasi-three level gain medium can be calculated with

$$G_{i-1} = \exp(g_{i-1}l), \tag{1}$$

with the small signal gain

$$g_{i-1} = \sigma_{eff,i-1} N, \qquad (2)$$

and the effective gain cross section

$$\sigma_{eff,i-1} = \beta_{i-1}(\sigma_{em} + \sigma_{abs}) - \sigma_{abs}, \qquad (3)$$

where $\beta_{i-1}$ represents the inverted fraction, $\sigma_{em}$ and $\sigma_{abs}$ the emission and absorption cross sections at the considered wavelengths, $N$ is the dopant ion density in the gain media and $l$ is the length of the gain medium.

After a single passage of an incoming fluence $J_{i-1}$, the fluence after the passage $J_i$ can be calculated with

$$J_i = J_{sat} T \ln[1 + G_{i-1} \exp(\frac{J_{i-1}}{J_{sat}}) - 1)], \qquad (4)$$

with the saturation fluence

$$J_{sat} = \frac{hc}{\lambda(\sigma_{abs} + \sigma_{em})}, \qquad (5)$$

and $T$ as the single pass transmission to generally account for losses in the system. To update the previous gain $g_{i-1}$ to the new average gain $g_i$, the following relation is used

$$g_i = \frac{J_{stor,i-1} - \left(\frac{J_i}{T} - J_{i-1}\right)}{J_{sat} l} = g_{i-1} - \frac{\frac{J_i}{T} - J_{i-1}}{J_{sat} l}, \qquad (6)$$

with the stored fluence

$$J_{stor,i-1} = J_{sat} \ln(G_{i-1}) = \sigma_{eff,i-1} N l J_{sat}. \qquad (7)$$

Rearranging Eq. (3) and Eq. (7), the new inverted fraction $\beta_i$ can be calculated by

$$\beta_i = \frac{\frac{J_{stor,i}}{J_{sat} l N} + \sigma_{abs}}{\sigma_{em} + \sigma_{abs}}. \qquad (8)$$

To account for the inversion decay losses during the pumping processes the inverted fraction is corrected by

$$\beta_i^* = \beta_i \exp\left(-\frac{\Delta t}{\tau_{gain}}\right), \qquad (9)$$

where $\tau_{gain}$ represents the upper state lifetime of the gain media and $\Delta t$ the duration of the fluence slice.

We define in the following the normalized gain $g_p$ and the gain depletion $g_{depl}$, which will be used to analyze the dynamics for gain build-up and gain-extraction during regenerative amplification. In the following, on the right-hand side of Eq. (10)-(12), we use the superscripted index [a] and [p] to indicate post-amplification and post-pumping values, respectively. We define the normalized gain $g_p$ as

$$g_p = \frac{J^p_{stor}(\lambda_a)}{J_{stor,PumpTransparency}(\lambda_a)} = \frac{\sigma^p_{eff}(\lambda_a)}{\sigma_{eff,PumpTransparency}(\lambda_a)}$$
$$= \frac{\beta^p(\sigma_{em}(\lambda_a) + \sigma_{abs}(\lambda_a)) - \sigma_{abs}(\lambda_a)}{(\beta_{PumpTransparency}\exp(-\Delta t/\tau_{gain}))(\sigma_{em}(\lambda_a) + \sigma_{abs}(\lambda_a)) - \sigma_{abs}(\lambda_a)}, \quad (10)$$

with

$$\beta_{PumpTransparency} = \frac{\sigma_{abs}(\lambda_p)}{\sigma_{em}(\lambda_p) + \sigma_{abs}(\lambda_p)}, \quad (11)$$

where $\sigma^p_g(\lambda_a)$ represents the effective gain cross section at the amplification wavelength $\lambda_a$, and $\beta^p$ is the inverted fraction after the pumping process. $\sigma_{g,PumpTransparency}(\lambda_a)$ represents the highest possible effective gain cross section at the amplification wavelength. In the frame of our FN based simulation model this happens for an inverted fraction $\beta_{PumpTransparency}$ at which the gain medium becomes transparent for the pump wavelength. A normalized gain of $g_p=1$ represents the maximum possible inverted fraction in the gain medium that can be achieved during a pumping process, and it equally represents the maximum possible stored fluence.

As a measure for the extracted fraction of that stored fluence after the amplification process, we define the gain depletion $g_{depl}$ as

$$g_{depl} = 1 - \frac{J^a_{stor}(\lambda_a)}{J^p_{stor}(\lambda_a)} = 1 - \frac{g^a(\lambda_a)}{g^p(\lambda_a)} = 1 - \frac{\sigma^a_{eff}(\lambda_a)}{\sigma^p_{eff}(\lambda_a)}$$
$$= 1 - \frac{\beta^a(\sigma_{em}(\lambda_a) + \sigma_{abs}(\lambda_a)) - \sigma_{abs}(\lambda_a)}{\beta^p(\sigma_{em}(\lambda_a) + \sigma_{abs}(\lambda_a)) - \sigma_{abs}(\lambda_a)}. \quad (12)$$

The gain depletion lies between 0 and 1, where 0 represents no energy extraction during pulse amplification and 1 represents a complete extraction of all stored energy.

To keep the numerical results as general and reproducible as possible we decided to choose an 'example system' for all numerical simulations in this paper. Consequently, where simulations and analogous measurements are shown in section 5, they demonstrate qualitative agreement and confirm tendencies that were predicted in the simulations. For the 'example Ho:YLF RA' we assume a beam diameter for the pump and cavity modes in the gain crystal of 1 mm. The gain crystal has a length of $l=10$ mm and the Holmium doping concentration is 1 % (Holmium dopant ion density $N=1.2\times10^{20}$ cm$^{-3}$ [20]). The seed fluence is 10 nJ/cm$^2$. The pump is assumed to be linearly polarized, by utilizing the absorption and emission cross sections at the pump and seed wavelengths for $\pi$-polarization from [21]. We take a value of 15 % for single pass losses (represented by $T=0.85$). An overview of the simulation parameters used for the simulations in this paper can be found in Table 2 in Appendix 2.

## 3. Operation regimes of regenerative amplifiers

This chapter numerically studies pulse amplification in four characteristic operation regimes for regenerative amplification as a function of the repetition rate, pump intensity and seed pulse fluence. The regimes were observed experimentally for example in [5,22], but there has been no study of these operation regimes as a function of the pump intensity, repetition rate and seed fluence.

Figure 2(a) illustrates four characteristic operation regimes (①–④) for regenerative amplification as a function of the repetition rate and pump intensity. The pump threshold (PT) indicates the pump intensity at which the amplifier starts providing gain to the seed pulses. Typical output pulse energies as a function of RT are sketched out for these regimes in Fig. 2(b-e). The average output pulse energy is indicated by solid red lines, and the optimum energy-extraction points in these 4 regimes are highlighted with the yellow stars. These stars indicate maximum pulse energy extraction while avoiding bifurcation or pulse instability. Regime ① allows a bifurcation-free operation, and stable pulses can be extracted at any RT. Although bifurcation appears for some RTs in regime ②, it is possible to extract high-energy output pulses at a single-pulsing RT beyond the bifurcation instability, as indicated by the yellow star. In regime ③, the yellow star shifts to an RT before the pulse instability. Finally, in regime ④, the highest average output pulse energy can be achieved before pulse instability appears. It is important to note that for regimes ①, ②, and ④, the yellow stars coincide with the peaks of the solid red lines. For the following, we define the bifurcation threshold (BT) as the repetition rate that separates regimes ① and ②.

Figure 3 and Fig. 4 present numerical results for the 4 operation regimes simulated with the 'example Ho:YLF RA'. To maintain generality, the pump intensity is given in multiples of PT and the repetition rate is normalized to the inverse excited state lifetime of Ho:YLF. Hence, a value of 1 represents a repetition rate of $1/\tau_{gain.}$. Fig. 3(a) is a direct representation of Fig. 2(a), evaluating the effect of the pump intensity on the operation regimes.

Fig. 3(b) presents simulation results for the inter-dependency of the pump intensity and the seed fluence on BT. For three pump intensities (1xPT, 5xPT and 15xPT), we observe that for an increase in the seed fluence by 6 orders of magnitude (from ~$10^{-3}$ nJ/cm$^2$ to 1000 nJ/cm$^2$), BT increases by a factor of ~ 1.5. Furthermore, for the complete range of seed fluences, BT increases by a factor of ~ 4.5 for an increase of the pump intensity from 1xPT to 5xPT (comparing black triangles and blue circles) and by a factor of ~ 13 for an increase from 1xPT to 15xPT (comparing black triangles and green stars).

The results demonstrate that the simplistic assumption that BT is solely decided by the inverse lifetime of the gain medium ($1/\tau_{gain}$) is only a limiting case for an RA with low pump intensity and low seed fluence. Based on the presented results, we propose the following proportionality between BT, the pump intensity and the seed fluence

$$BT \propto \frac{1}{\tau_{gain}} \cdot I_{pump} \cdot J_{seed}^b, \qquad (13)$$

with $I_{pump}$ as the pump intensity, $J_{seed}$ as the seed pule fluence and $b$ as a fitting parameter. The relation suggests a linear dependence of BT with the pump intensity and a dependence on the seed fluence that follows a power-law. To verify this assumption, we fitted the simulation results from Fig. 3(b) with

$$BT \cdot \tau_{gain} = f_1(I_{pump}) \cdot f_2(J_{seed}^b), \qquad (14)$$

where $f_1$ and $f_2$ represent the dependency on the pump intensity and the seed fluence with

$$\begin{aligned} f_1 &= m \cdot (\frac{I_{pump}}{PT} - 1) + n, \\ f_2 &= c + d \cdot J_{seed}^b, \end{aligned} \qquad (15)$$

with *m, n, c, d* and *b* as fitting parameters. The fitted BT curves are also shown in Fig. 3(b) with the solid lines. The fitted parameter values for *m, n, c, d* and *b* are listed in Table 1. We find good agreement for seed fluences that are much smaller than the total stored fluence in the gain medium, in this particular case, for seed fluences < 1000 nJ/cm$^2$.

Figure 3(c) presents the operation regimes ①–④ as a function of the normalized repetition rate and the seed fluence, for a pump intensity of 15xPT. The BT-line from Fig. 3(b) is re-plotted, now for a larger range of seed fluences and normalized repetition rates. We observe that for seed fluences > 1000 nJ/cm$^2$ the simulated results (green hollow stars) and the fitted curve (orange solid line) for BT start to deviate. For seed fluences exceeding a certain cut-off value (here $10^6$ nJ/cm$^2$), no pulse instability appears anymore, independent of the repetition rate. These results more generally expressed in the total amplifier gain, we observe good agreement for gains of > $10^6$, the curves start to deviate for gains between $10^3$ - $10^6$, and the cut-off appears for gains ≲ $10^3$.

Table 1. Parameter values for Eq. (14) – (15), fitted to the simulation results for BT in Fig. 3(b).

| Parameter for $f_1$ | Parameter for $f_2$ |
|---|---|
| m = 0.5 | c = 1.29 |
| n = 0.57 | d = 4.53 |
|  | b = 0.12 |

The horizontal orange dotted line in Fig. 3(a) and 3(d) corresponds to a pump intensity of 15xPT. For operation along these lines, we consider in Fig. 4 the highest-energy extraction point at a single-pulsing RT (indicated by the yellow stars in Fig. 2). At this extraction point, the output pulse fluence and the average output pulse intensity

(output pulse fluence × repetition rate) are plotted as a function of the normalized repetition rate in Fig. 4(a). Figure 4(b) shows the corresponding normalized gain $g_p$ and the gain depletion $g_{depl}$. The thin vertical dotted lines separate regimes ① to ④.

Regime ① is characterized by a saturated normalized gain (blue solid line) and high gain depletion (green dotted line), as shown in Fig. 4(b). This means that a high fluence is stored in the gain medium during the pumping process and most of it is depleted in the following amplification process. Consequently, high output fluences (black dashed-dotted line) can be extracted [Fig. 4(a)]. Within regime ①, the output fluence is saturated and remains nearly constant with an increase in the repetition rate, while the average output intensity (red dashed line) increases almost linearly.

In regime ②, the pump intensity is no longer sufficient to saturate the normalized gain during the pumping processes and, consequently, the output fluence decreases (black dashed-dotted line). Here we observe the onset of bifurcation. The peak in the average output intensity (red dashed line) may represent the 'best' compromise between high output fluence extraction and high average output intensity. The transition between regimes ② and ③ is marked by a sudden step in the normalized gain (blue solid line) and gain depletion (green dotted line). This corresponds to a jump of the yellow star from after-pulse instability in Fig. 2(c) to before-pulse instability in Fig. 2(d). Consequently, less gain is depleted and the normalized gain after the successive pumping processes increases.

The operation regimes ③ and ④ are characterized by low gain depletion and consequently low output fluence, as only a small amount of the stored fluence is depleted. We observe that the output fluence (black dashed-dotted line) decreases almost logarithmically with the repetition rate. At the beginning of regime ③, the average output intensity (red dashed line) is low but increases with an increase in the repetition rate. This is caused by the shift of the yellow star in Fig. 2(d) towards the peak of the average pulse energy [red line in Fig. 2(e)] as the repetition rate increases. In regime ④, the normalized gain (blue solid line) stays nearly constant, while the average output intensity (red dashed line) reaches its saturated maximum.

## 4. Gain dynamics in regenerative amplifiers

The focus in this chapter lies on exposing the actual gain dynamics during RA operation. The simulations in the previous section were conducted without pump noise and, hence, demonstrated the inherent susceptibility of an RA to show pulse instability. In contrast, in this section the simulations were conducted with pump noise. The general RA gain dynamics are not affected by this rather small-scale pump intensity fluctuation. However, only by including pump noise we will observe the emergence of a highly stable operation point.

A pump noise of 3 % was chosen in the simulations, implemented in accordance to the definition at the beginning of section 2. An equivalent definition was used to analyze the simulation results in terms of the output noise.

Figure 5(a) presents the output noise simulated for the vertical pink line in Fig. 3(a), drawn as a function of the RT and pump intensity. We observe regimes with high pulse instability (yellow and red), as well as regimes without strong pulse instability (green and blue). Analysis of the simulation results in terms of normalized gain $g_p$ and gain depletion $g_{depl}$, as defined in Eq. (10)-(12), highlight the underlying gain dynamics. Figure 5(b) presents the re-drawn simulations results from Fig. 5(a) as a function of $g_p$ and $g_{depl}$.

To elucidate the mapping from Fig. 5(a) to 5(b), three vertical and three horizontal dotted lines are chosen in Fig. 5(a), marked as (1)–(6). These are re-drawn in Fig. 5(c) as trajectories in the normalized gain and gain depletion space. The trajectories demonstrate the RA gain dynamics when either the RT is changed for a fixed pump power [curves (1)–(3)], or alternatively, when the pump power is changed for a fixed RT [curves (4)–(6)]. The trajectories (1), (2), and (3) represent examples for operation in the regimes ①, ②, and ③. The trajectories (1)–(3) start with a gain depletion of 0 and increase to higher gain depletion with an increase in RT. Following the curves (1)–(3), the increase in gain depletion is accompanied by a decrease in normalized gain because the pumping process cannot restore the normalized gain to its initial value anymore.

In the white area in Fig. 5(c), the simulation did not produce values. The ranges producing stable output pulses are characterized either by low gain depletion (very little of the stored fluence extracted from the gain medium) or by high gain depletion (most of the stored fluence extracted).

It is important to note that on including of pump noise to the simulations, we observe in Fig. 5(c) the emergence of a highly stable operation point (deep blue) at a specific gain depletion value (in this case, $g_{depl} \sim 0.85$). The exact position mainly depends on the cavity losses and pump intensity, which will be discussed in Section 5.

This highly stable operation point can only be accessed if the pump intensities are sufficiently high to drive the RA into the operation regime ① (for approx. $g_p > 0.85$) or regime ② (approx. $0.75 < g_p < 0.85$).

Figure 5(d) demonstrates an analogous analysis for the gain dynamics in regime ④ (simulated at a normalized repetition rate of 750). Here, operation is characterized by rather low gain depletion values and, consequently, low output pulse fluences. In this regime, higher gain depletion does not necessarily coincide with higher output fluence. Once the RA reaches the maximum output fluence, an increase in gain depletion (potential increase in output fluence) for example by increasing RT, is overcompensated by the attendant decrease in the normalized gain and stored fluence.

Figure 6 demonstrates the noise and gain dynamics for three different seed fluences. The parameters are chosen identical to the simulations in Fig. 5(c), however to demonstrate the isolated effect of the seed fluence on the gain dynamics, we conducted the simulations without losses and without pump noise. The spiky appearance of the border between the stable and the unstable region is an effect of the RT that can only be varied in the simulations in multiple integers of a single RT. The graphs show that for an increase of the seed fluence, the unstable region shrinks, and consequently, stable and bifurcation-free operation is possible for a larger parameter space. We observe that when pump noise and losses are included in the simulations, a highly stable operation point appears again at the same position as in Fig. 5(b), independent of the seed fluence.

## 5. RA noise optimization and low-noise operation

This section presents both numerical and experimental results for the output noise of an RA that is operated at high normalized gain and gain depletion values close to a highly stable operation point [located in Fig. 5(b)]. In section 5.1, RA operation is studied numerically as a function of the RA parameters RT, single pass losses, pump intensity and seed fluence. In section 5.2, laboratory measurements for operation at analogous highly stable operation points are presented as a function of the parameters pump intensity, seed fluence, crystal holder temperature and RT.

It needs to be noted that although the numerical 'example Ho:YLF RA' and laboratory Ho:YLF RA setup were both operated in operation regime ①, the operational parameters of the two systems are different. Consequently, the measurements intend to demonstrate qualitative agreement and to confirm tendencies that were predicted in the simulations.

*5.1 Simulations*

All simulations conducted in this section can be localized in Fig. 3(a) along the vertical pink dashed-dotted line (or at points on that line), representing operation at a normalized repetition rate of 15. As a measure for the output noise, we use as before the standard deviation of the output pulse fluence in percent of the mean output fluence (for the simulations), or an equivalent expression in terms of energy (for the measurements). Figure 7(a) presents the simulated output noise as a function of RT for three different combinations of pump and seed noise. The pump intensity was chosen for operation in regime ① (22xPT). Two ranges of operation can be distinguished:

(i) For up to RT=13, the gain is not yet depleted and the pulse energy increases with an increase in RT. Here, both the pump and the seed noise contribute to the output noise.

(ii) Once the RA approaches the maximum pulse energy, the effect of the seed noise [green curve in Fig. 7(a)] becomes small compared to the effect of the pump noise [black curve in Fig. 7(a)]. In this range, the pump noise completely dominates the output noise, i.e. it becomes almost identical to that for the RA exhibiting both pump and seed noise [black and red line in Fig. 7(a)].

For the local maxima (RT=15) and minima (RT=18) of the output noise in Fig. 7(a), the output noise was found to be linearly dependent on the input pump noise, as shown in Fig. 7(b).

Figure 7(a) and 7(b) emphasize that to realize an RA that produces a highly stable output pulse energy, there is a need for first, the identification of a low-noise operation point and, second, a low-noise pump source. Furthermore, for the extraction of pulse energies close to gain depletion, a low-noise seed source is of less significance than a low-noise pump source.

In the following we focus on the noise characteristics in the gain-depleted range close to a noise minimum. As seed noise can be neglected in this range, all simulations were conducted with pump noise only. Because of the linear dependence of the output noise and the input pump noise [shown in Fig. 7(b)], in the following we normalized all simulation results for output noise to the input pump noise. Thus, a normalized output noise of 1 corresponds to an equal input pump and output noise.

Figure 8(a) presents the physical origin of the local noise minima. Figure 8(a) shows simulation results with the 'example Ho:YLF RA' for RT=19 as a function of the pump intensity for the four different single pass losses 0 %, 5 %, 10 %, and 15 %. For a system without losses, no intermediate minimum is present. For increasing losses, the noise minimum becomes more and more pronounced. Consequently, the output pulse fluence increases with a decrease in the losses.

The simulations are conducted as a function of the pump intensity and not as a function of RT. In terms of accessing a noise minimum, both parameters can be used, but only the pump intensity can be varied in sufficiently small steps to resolve the noise fine structure, as will be further emphasized in Fig. 10(a) and Fig. 10(b).

By plotting the noise curves as a function of the gain depletion one can see that the noise minima are located at different gain depletion values [Fig. 8(b)], at higher depletions for lower losses. As mentioned before, the curve representing a loss-less system lacks an intermediate minimum.

The output fluence in Fig. 8(b) was normalized to the total fluence that is stored in the gain material after the pumping cycle with

$$J_{norm} = \frac{J_{out}}{J_{stor}}. \tag{16}$$

The interpretation is straightforward. For the loss-less system, the normalized output fluence increases linearly with the gain depletion and reaches a value of 1 for a gain depletion of 1. In this case, the normalized output fluence is a direct representation of the gain depletion. On introducing losses, this direct representation is not valid anymore. In this specific case, simulating single pass losses of 5 %, 10 %, and 15 % leads to a maximum normalized output fluence of 0.69, 0.51, and 0.39, respectively. This means that although most of the stored energy is extracted (represented by a high gain depletion), a significant fraction of the stored energy is lost and dissipated into the environment.

Figure 9 further investigates the accessibility and characteristics of these highly stable operation points. Fig. 9(a) and Fig. 9(c) show the normalized output noise and the output fluence as a function of the pump intensity and the seed fluence, respectively. These curves are re-plotted in Fig. 9(b) [re-plot of Fig. 9(a)] and 9(d) [re-plot of Fig. 9c)] as a function of gain depletion. The output was simulated for three different RTs. Fig. 9(a) and Fig. 9(c) show that for a decrease in the RT, the noise minima shift to a higher pump intensity or higher seed pulse fluence respectively. This is also reflected in the shift of the noise minima to higher gain depletion in Fig. 9(b) and 9(d).

For operation at the noise minima, we observe in Fig. 9(a) for an increase in the pump intensity [and equivalently for an increase in the gain depletion in Fig. 9(b)] an increase in the output fluence [Fig. 9(a)] as well as in the normalized output fluence [Fig. 9(b)]. We assume that the main reasons for the increase in the output fluence and in the normalized output fluence are an increase in the stored fluence in the gain medium (due to the higher pump intensity) and a decrease in the accumulated losses (due to the lower RT). In the following, we try to quantify these reasons by comparing to output fluence in Fig. 9(a) and the normalized output fluence in Fig. 9(b). While the output fluence is affected by the pump intensity and the accumulated losses, the normalized output fluence is mainly affected by the accumulated losses alone (due to its normalization to the stored fluence in the gain medium). For operation at a noise minimum, we observe for this specific case that the increase in the normalized output fluence [in Fig. 9(b)] is roughly half the increase in the output fluence [in Fig. 9(a)]. Therefore, we conclude that for this specific RA, the increase in the output fluence is influenced equally by the increase in the stored fluence and the decrease in the accumulated losses.

We observe in Fig. 9(a) that for an increase in the pump intensity, the RA can produce output pulses with a higher stability (decreased output noise).

Fig. 9(c) and 9(d) show the normalized output noise as a function of the seed fluence [Fig.9(c)] and gain depletion [Fig. 9d)]. Similar to the results in Fig. 9(a), increasing the RT shifts the noise minima to a decreased seed pulse fluence.

Rather counter-intuitively, when operating at RT=17 instead of RT=15, the output fluence and the normalized output fluence both increase by approximately 1.5 % despite more than an order of magnitude less seed fluence (at the corresponding noise minimum). This is rather unexpected, as one would assume that an increase in RT leads to a decrease in the output pulse fluence due to an increase in accumulated losses.

However, we believe that this observation is valid as long as the seed fluence is small compared to the total stored fluence in the gain medium and other loss channels (such as ASE or spontaneous inversion decay during pulse amplification) are negligible. For high-gain RAs with long-lifetime gain materials, such as for Ho:YLF, this precondition is fulfilled in most cases. The above observation can be made traceable as follows:

(i) Considering a loss-less system, the stored fluence in the gain medium represents the maximum output pulse fluence that can be extracted. Unlike higher pump intensities, higher seed fluences do not change the stored fluence in the gain medium.

(ii) When losses are included, the extractable output fluence is lowered by the accumulated losses. The simulations suggest, however, that for operation close to a noise minimum, a lower or higher seed fluence (and consequently, more or less RTs) do not change the absolute loss. This can be seen in Fig. 9(d), where the lines for the normalized output fluence for the different RTs overlap each other.

(iii) From Fig. 9(d), it can be seen for higher RTs that the output fluence at the corresponding noise minimum shifts closer to the maximum of the output fluence curve. This leads to an increase in the output fluence for lower seed energy if the RA is operated at the corresponding noise minimum. This behavior agrees with measurements as well, which are discussed in Fig. 10 (c).

*5.2 Measurements*

We operated our Ho:YLF RA [6] in regime ① (at a repetition rate of 100 Hz) and measured the output noise and the output pulse energy as a function of various parameters. The details of the RA setup can be found in the Appendix 1.

The measurements in dependence of the pump power in Fig. 10(a) and the seed energy in 10(c) are analogous to the simulations in Fig. 9(a) and 9(c), respectively. The measurements in Fig. 10(b) and 10(d) have no analogous part in the previous simulations-section but add further information that we consider practically relevant for the operation and noise optimization of an Ho:YLF RA.

The RA was seeded with pulses from an Ho:fiber oscillator [23] that produced pulses centered around 2053 nm with a bandwidth of 7 nm. The pulses were stretched with a chirped volume Bragg grating (CVBG) prior seeding. The RA was pumped with a commercial CW Tm:fiber laser that produces up to 120 W of randomly polarized light. Unfortunately, the output power of this laser could not be tuned in fine steps. The measurements presented in Fig. 10(a) and 10(b), however, required a fine tuning of the pump power. This was realized by a polarization of the pump laser output and by usage of a further polarizer and λ/2-waveplate. The measurements that did not require a variation of the pump power, in Fig. 10(c) and 10(d), were conducted with 18 W of the unpolarized pump laser output. To avoid water condensation at low crystal temperatures, the complete RA set-up was purged with a constant flow of dry nitrogen, keeping the humidity level < 2 %. We observed that purging the RA cavity also significantly improved the stability of the amplified output pulses, related to an improved pump laser stability under nitrogen atmosphere [see Appedix 2 in Fig. 11(e)]. If not specified otherwise, the RA was seeded with pulse energies of 690 pJ and the temperature of the thermo-electrically controlled Ho:YLF crystal holder was kept at 18 °C.

The presented pulse energies correspond to the uncompressed output, measured with a commercial calibrated energy meter. For each measurement, consecutive output pulses were recorded for 30 seconds and analyzed with respect to pulse energy and output noise.

Figure 10(a) presents the measured output noise and output pulse energy as a function of the pump power for RT=12 an d RT=14. The minimum noise values are 0.76 % and 0.95 % for 12 and 14 RTs respectively. As predicted by the simulations in Fig. 9(a), for an increase in the pump power, the amplified output pulses can be coupled out at an earlier RT. At the corresponding noise minima, the output noise decreases and the output pulse energy increases for an increase in the pump power.

In this particular case, the stability was improved by ∼ 20 % and the pulse energy increased by ∼ 50%. The range of pump powers at which a minimum noise can be achieved is quite narrow. For instance, on reduction of the pump power by only 1.3 % (from 15.7 W to 15.5 W), the noise minimum increased by ∼ 200 %, from 0.76 % to 1.49 %. Therefore, if the pump power is not fine-tuned, it is easily possible to miss the lowest noise operation point. This is emphasized in Fig. 10(b), which presents the output noise and output energy as a function of RT for three different pump powers. The pump powers of 15.7 W and 14.1 W at RT=12 and RT=14 respectively represent the two noise minima in Fig. 10(a). The curve for the pump power of 15.5 W represents operation with a pump power slightly detuned by 0.2 W from the optimum pump power, effecting a twofold increase in the output noise at RT=12. Operation at any other RT leads to higher output noise, thereby missing the highest stable operation point.

Fig. 10(c) presents the measurements to study the effect of the seed pulse energy on the output noise and the output energy for different RTs. As suggested by the simulations in Fig. 9(c), we observe a slightly lower noise minimum for an increase in the seed energy, at a lower RT. Furthermore, the output energy at the corresponding noise minima shifts closer to the maximum of the corresponding output pulse energy curve. This leads to the counter-intuitive effect that at a noise minimum, the output pulse energy increases despite lower seed energy. In particular, we measured noise values of 0.24 % and 0.27 % for the noise minima at seed pulse energies of 280 pJ and 19 pJ, with corresponding output pulse energies of 2.45 mJ and 2.5 mJ, respectively. In this case, operation at the higher RT results in a stability improvement of 11 %. At the same time, despite more than an order of magnitude more seed energy, the amplified output energy decreases by 2 %.

During operation with the Ho:YLF RA, we experimentally discovered that a variation of the Ho:YLF crystal holder temperature allows a noise optimization similar to the optimization descripted in dependence of the pump power and seed energy. Figure 10(d) shows the output noise and the pulse energy as a function of the crystal holder temperature for RT=11 to RT=14. Similar to the previous noise curves [Fig. 10(a-c)], there is a distinct noise minimum, in this case at a specific crystal holder temperature. For a decrease in the crystal holder temperature, we also observe a significant increase in the output pulse energy. With a decrease in the crystal holder temperature from 18 °C to -16 °C, the output pulse energy increases almost linearly by a factor of 2.6, from 2.5 mJ to 6.43 mJ. Currently, the temperature dependence of the emission and absorption cross-sections is not implemented in the simulation model. However, based on the observations and findings described in this paper, we propose the following physical interpretation. We observe that with a decrease in the Ho:YLF crystal temperature, the seed pulses experience a stronger amplification, caused by higher absorption cross sections at the pump wavelength and higher emission cross sections at the seed wavelength at lower temperatures [9,24]. Consequently, the optimum gain depletion is reached with fewer RTs and the pulses can be coupled out earlier. Hence, the output energy increases at lower temperatures due to fewer accumulated losses.

We observed that the minimum noise values measured with linearly polarized light were higher by up to a factor of 3 than the ones measured with randomly polarized light. These noise values, however, cannot be directly compared as the actual noise from the pump laser might be different for the two cases. Two circumstances may be expected to contribute to the increased noise observed:

(i) The pump laser was operated at 32 W, almost twice the value needed for the measurements with random polarization at 18 W. Therefore, the higher noise level of the RA output could be explained with different noise levels of the pump.

(ii) The fraction of the polarized light that is used to pump the RA fluctuates, thus introducing an additional source of pump noise.

Figure 10(e-g) shows recorded oscilloscope traces for the intra-cavity pulse build-up typical for operation at a noise minimum and for operation before and after the minimum (with a less or more depleted gain) respectively. These operation points were accessed by controlling the crystal holder temperature while keeping RT fixed. We observe that for T=T1 [in Fig. 10(e)] the output pulses exhibit the lowest noise, whereas the pulses exhibit higher noise for T < T1 and T > T1 [in Fig. 10(f) and 10(g)].

## 6. Conclusions

We presented a comprehensive analysis of pulse amplification in Ho:YLF RAs. Although this study was conducted with an Ho:YLF RA, the findings are expected to be more generic and lead to a more complete understanding of RA operation regimes, gain dynamics and low-noise operation.

In Section 2, we presented a numerical simulation method, based on the iterative Frantz-Nodvik formalism, simulating consecutive pumping and amplification cycles. The model also accounts for noise originating from the pump and seed sources. The model is computationally fast, which is important for simulations involving a large number of consecutive pumping and amplification cycles. This is crucial for a statistical analysis of results when RA parameters are varied finely or over a large range.

In Section 3, we numerically identified RA operation regimes with respect to the repetition rate and pump intensity. We recognized four different operation regimes that show different output pulse characteristics. In regime ①, for low repetition rates, no bifurcation instability exists. In regime ②, bifurcation appears, but the highest average pulse energy is at a single-pulsing RT beyond the bifurcation instability. In regime ③, the highest output energy at a single-pulsing RT is limited due to the onset of pulse instability. In regime ④, the highest average pulse energy can be extracted at an RT before pulse instability appears, and consequently the instability can practically be ignored under such circumstances. Our simulations showed that the repetition rate separating regimes ① and ② shifts to higher repetition rates for increased pump intensities and higher seed fluences. The results indicate that the simplistic assumption that the onset of bifurcation is solely decided by the inverse lifetime of the gain medium ($1/\tau_{gain}$) is only a limiting case for an RA with low pump intensity and low seed fluence. We empirically find for high gain RAs with gains > $10^6$, a linear proportionality between BT and the pump intensity, and a proportionality following a power-law for the seed fluence.

In Section 4, we presented a numerical analysis of the gain dynamics in RAs operating in regimes ①–④ with respect to gain build-up and gain depletion in the gain medium. On including noise in the simulations, we identified an operation point that allows the extraction of highly stable output pulses. This operation point corresponds to a high gain build-up during RA pumping and a high gain depletion during pulse amplification. We also showed that for an increase in the seed energy, bifurcation-free operation is possible for a wider parameter range.

In Section 5, we studied both experimentally and numerically the output noise and output pulse energy at highly stable operation points. The fundamental cause for the observed noise minima was numerically identified to be the RA losses. This means for every real laboratory RA that is operated in the regimes ① or ② that there always is a noise minimum located in close proximity to the highest output pulse energy. Our findings strongly suggest that any RA parameter that allows a fine-adjustment of the gain depletion can be used to access and optimize these highly stable operation points.

To experimentally study analogous highly stable operation points with an Ho:YLF RA, we varied the pump intensity, seed fluence, number of round trips and crystal-holder temperature and we found good qualitative agreement between the simulation results and the measurements. In this specific case, optimization of the RA parameters led to highly stable output pulses from our Ho:YLF RA with measured pulse energy fluctuations of only

0.19 % (standard deviation). The noise curves show very characteristic noise minima that can be quite narrow (noise increases by 100 % for a change in the pump power of only ∼ 1.3 %). Therefore, the lowest-noise operation points can be easily missed if the RA operational parameters are chosen as slightly different from their optimum values, or if they drift with time and cause the RA output noise to move up and down the corresponding noise curves.

**Funding**

This work has been funded through the Max-Planck-Gesellschaft (Max-Planck Society).

**Acknowledgments**

The authors thank Haider Zia for helpful discussions.

**Appendix 1: The Ho:YLF RA setup**

Figure 11(a)-(d) presents details of the Ho:YLF RA setup [6] that was used for the measurements in 5.2. The oscillator pulses [23] to seed the RA were stretched with a CVBG in a double-pass configuration (not shown in the figure). After stretching, the pulse energy was 690 pJ with a spectral bandwidth of 3.25 nm. Based on the specified chirp rate of 41 nm/ps for the CVBG, we estimated the pulses to be chirped to a pulse duration of 267 ps.

The pump chain, shown in Fig. 11(a), consists of two telescopes and an optional polarization splitting attenuator. The simulated path of rays is shown in Fig. 11(d). The telescopes image the end of the pump fiber into a 4 cm long Ho:YLF crystal that has a Holmium doping concentration of 1 %. The image diameter in the Ho:YLF crystal is ∼ 1 mm and can be adjusted by simply moving the lenses F3 and F4 of Telescope 2.

Due to the recent commercial availability of high power and large aperture Faraday rotators for the wavelength of 2 μm (FastPulse Technology, Inc.), the RA cavity is arranged linearly [shown in Fig. 11(b)]. A typical output beam shape, with a beam diameter of ∼ 3.1 mm, is also shown in Fig. 11(b). It was measured in a distance of about 2.5 m after the RA output port and after being compressed with a chirped volume Bragg grating (CVBG) (not shown in the sketch). In this specific case, the uncompressed output pulse energy was 9 mJ at a repetition rate of 100 Hz. A randomly polarized pump power of ∼ 24 W was used. The simulated mode in the Ho:YLF crystal has a radius of 475 μm, resulting in a peak pulse fluence in the crystal of ∼ 2.5 J/cm$^2$. We did not observe crystal damage for these parameters.

During RA operation, the setup was constantly purged with nitrogen to avoid condensation of water on the Ho:YLF crystal and because the output pulse stability of the RA was observed to be significantly improved under nitrogen atmosphere. This improved stability can be connected to a significantly more stable and improved pump beam. A typical pump spot in the Ho:YLF crystal plane under laboratory and under nitrogen atmosphere is shown in Fig. 11(e). The disturbed pump beam shape might be attributed to the water vapor in the laboratory atmosphere that has absorption lines close to the lasing wavelength of the pump laser at 1940 nm.

**Appendix 2: Overview of simulation parameters**

All simulations were conducted for an 'example Ho:YLF RA' with the general parameters as follows: The beam diameter for the pump and cavity modes in the gain crystal is 1 mm, the gain crystal has a length of $l=10$ mm and the Holmium doping concentration is 1% (Holmium dopant ion density $N=1.2\times10^{20}$ cm$^{-3}$ [20]). The pump is assumed to be linearly polarized, by utilizing the absorption and emission cross sections at the pump and seed wavelengths for $\pi$-polarization from [21]. The other parameters were chosen in accordance to Table 2. It needs to be mentioned that for a better visibility, the figures often only show an excerpt of the complete simulated range.

**Table 2. Parameters for the simulations presented in this paper.**

| Figure | Pump power | | Repetition rate | | Seed fluence | Pump noise | Single pass transmission |
|---|---|---|---|---|---|---|---|
| | W | multiples of PT (1×PT = 4.5 W) | kHz | Normalized (multiples of 1/$\tau_{gain}$) | nJ/cm$^2$ | % | |
| **Fig. 3** | | | | | | | |
| (a) | 0-250 | 0-55 | 0-100 | 0-1500 | 10 | 0 | 0.85 |
| (b) | 4.5/22.5/67.5 | 1/5/15 | 0.03-2 | 0.5-30 | 10$^{-3}$-10$^3$ | 0 | 0.85 |
| (c) | 67.5 | 15 | 0.01-10 | 0.15-150 | 10$^{-3}$-10$^9$ | 0 | 0.85 |
| **Fig. 4** | 67.5 | 15 | 0.01-100 | 0.15-1500 | 10 | 0 | 0.85 |
| **Fig. 5** | | | | | | | |
| (a-c) | 0-250 | 0-55 | 1 | 15 | 10 | 3 | 0.85 |
| (d) | 0-250 | 0-55 | 50 | 750 | 10 | 3 | 0.85 |
| **Fig. 6** | | | | | | | |
| (a) | 0-250 | 0-55 | 1 | 15 | 10 | 0 | 1 |
| (b) | 0-250 | 0-55 | 1 | 15 | 10$^3$ | 0 | 1 |
| (c) | 0-250 | 0-55 | 1 | 15 | 10$^5$ | 0 | 1 |
| **Fig. 7** | | | | | | | |
| (a) | 100 | 22 | 1 | 15 | 10 | 0-1 | 0.85 |
| (b) | 100 | 22 | 1 | 15 | 10 | 0-5 | 0.85 |
| **Fig. 8** | 50-300 | 11-67 | 1 | 15 | 10 | 3 | 0.85\0.9\0.95\1 |
| **Fig. 9** | | | | | | | |
| (a) | 50-300 | 11-67 | 1 | 15 | 10 | 3 | 0.85 |
| (b) | 100 | 22 | 1 | 15 | 1-10$^3$ | 3 | 0.85 |

**Figures**:

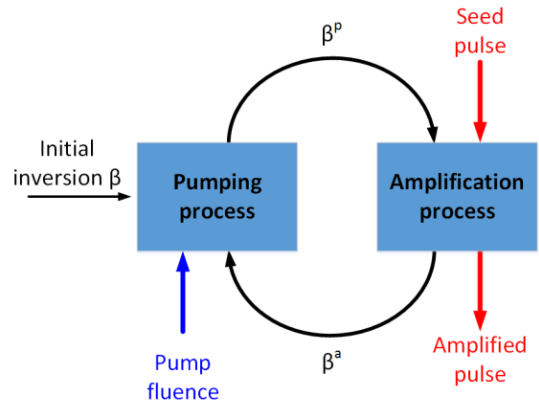

Fig. 1. Simulation of consecutive pumping and amplification cycles. The inverted fraction $\beta^p$ and $\beta^a$ from the pumping or amplification cycle is fed to the following amplification or pumping cycle, respectively.

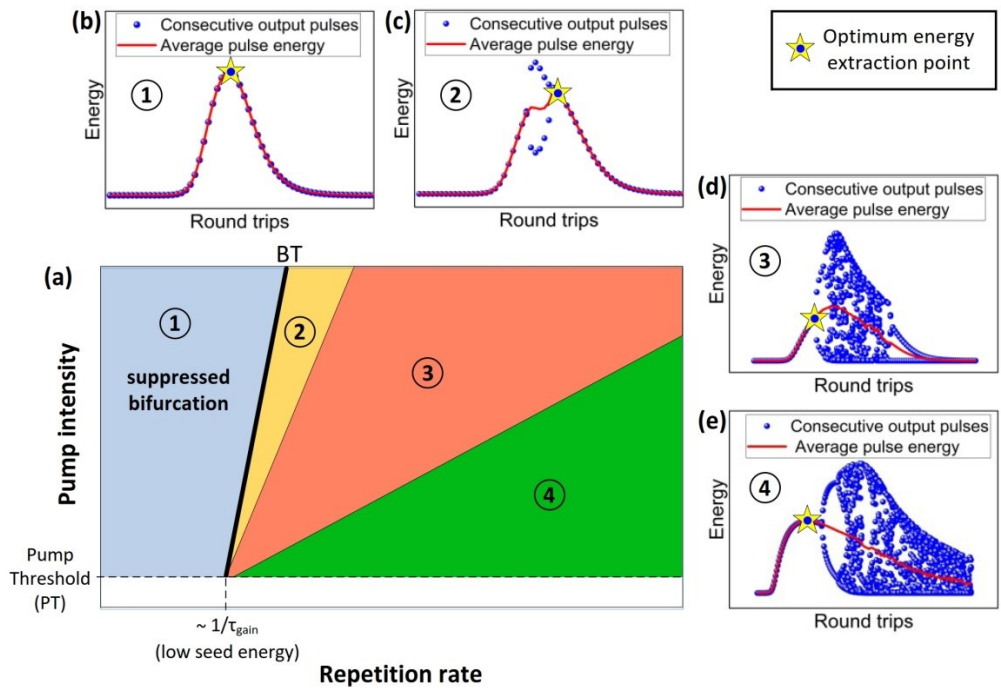

Fig. 2. (a) Illustration of RA operation regimes as a function of the repetition rate and pump intensity. (b-e) Typical output pulse energies for operation in these regimes are sketched out as a function of RT. The yellow starts indicate the optimum energy extraction points. Regime ① represents a bifurcation-free regime. In regime ②, high pulse energies can be extracted by operation beyond the bifurcation instability. In regime ③, the maximum pulse energy at a single-pulsing RT is limited by the onset of the pulse instability. In regime ④, the maximum pulse energy at a single-pulsing RT can be extracted before the pulse instability. The bifurcation threshold (BT) separating regimes ① and ② coincides with $\sim 1/\tau_{gain}$ only for low pump intensities and low seed energies.

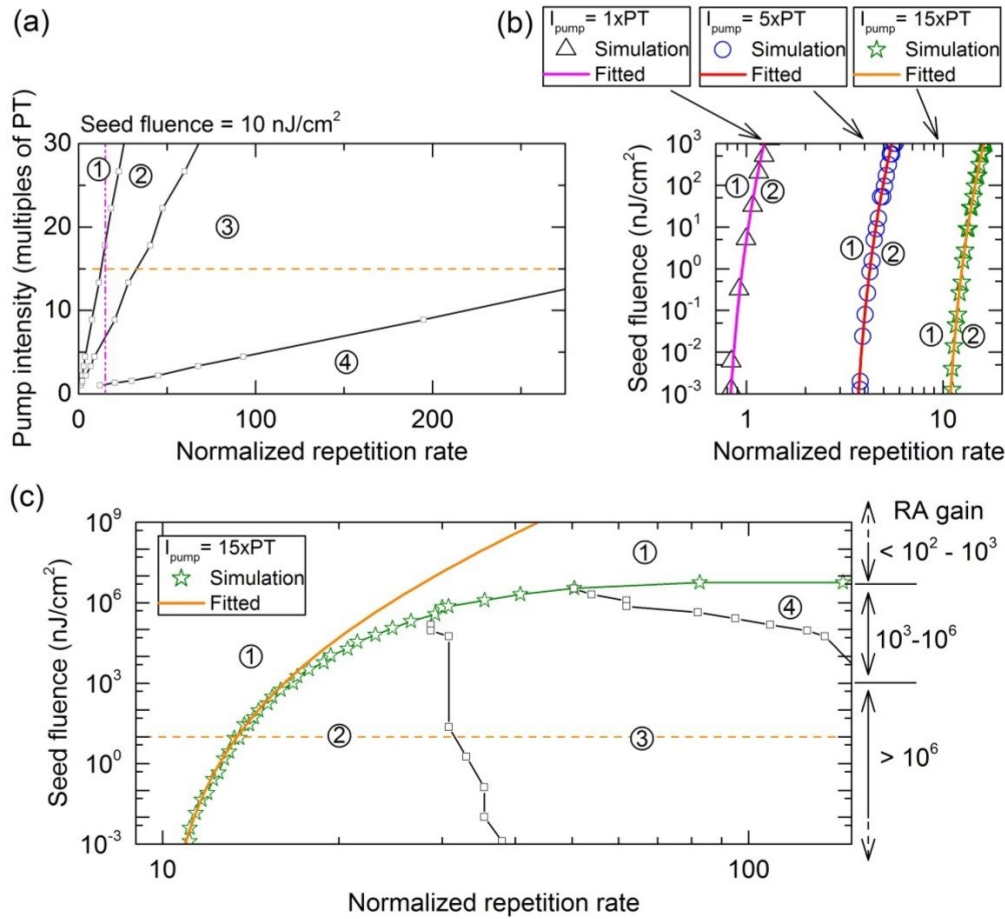

Fig. 3. (a) Simulation results for the separation lines of the RA operation regimes ①-④, as a function of the pump intensity. In Fig. 3(b), the BT is shown for small seed fluences (<< stored fluence in the gain medium) for the pump intensities 1xPT, 5xPT and 15xPT. (c) Operation regimes in dependence of the seed fluence, for a pump intensity of 15xPT. The solid lines in pink, red and orange in Fig. 3(b) and in orange in Fig. 3(c) represent the fitted BT with Eq. (14)-(15). The used parameter values are listed in Table 1. Good agreement between the simulation results and fitted curves for BT can be observed for low seed fluences (high gains). The horizontal orange dotted line-cuts in (a) and (c) represent a pump intensity of 15xPT, for which output results are presented in Fig.4. The vertical pink dash-dotted line represents a normalized repetition rate of 15, along which the gain dynamics will be studied in section 4 [Fig. 5(a-c)].

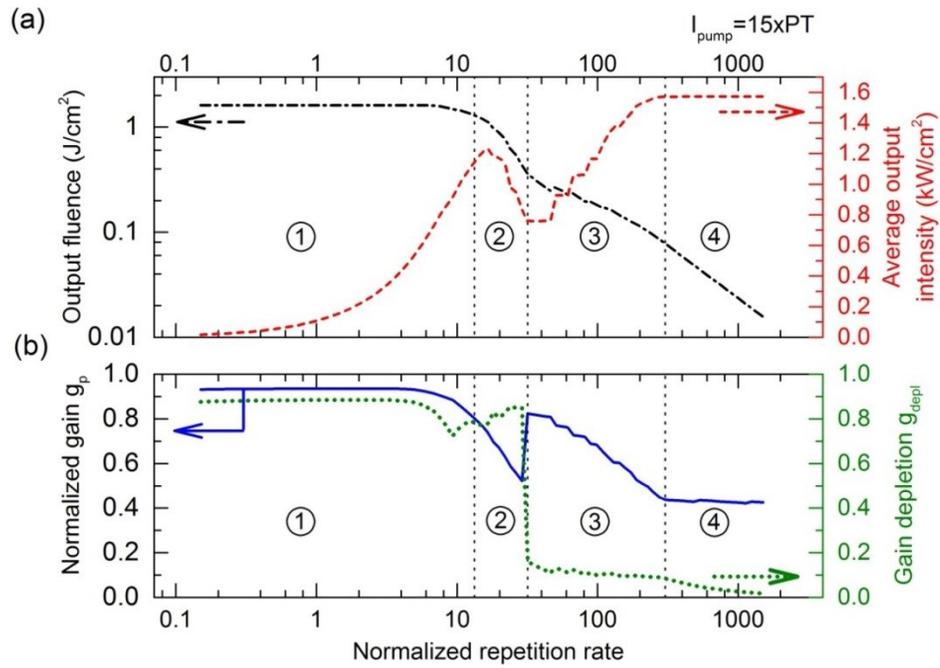

Fig. 4. Cut through Fig. 3(a) at a pump intensity of 15xPT. As a function of the repetition rate, (a) shows the output pulse fluence and the average output intensity, and Fig. 4(b) shows the normalized gain and the gain depletion. The thin dashed vertical lines represent the border lines between the different operation regimes. The values correspond to the optimum energy extraction points [indicated by yellow stars in Fig. 2(b-e)].

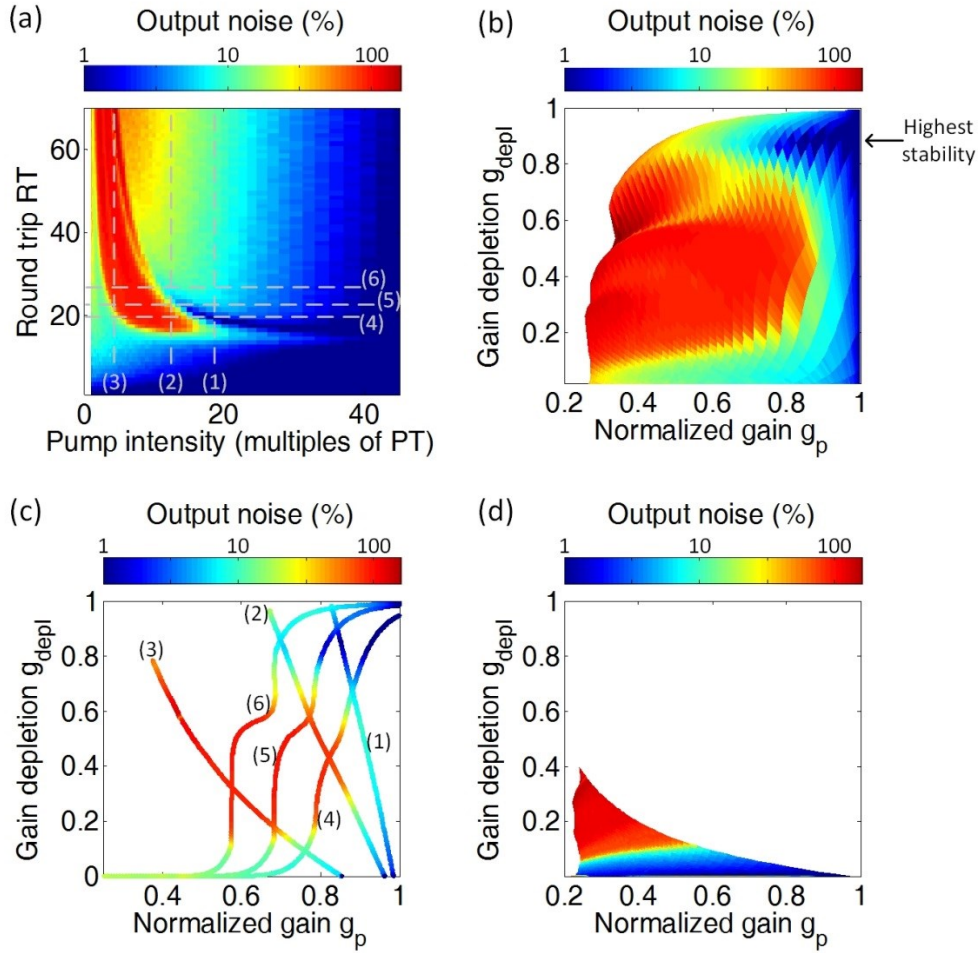

Fig. 5. (a) Output noise as a function of the pump intensity and the RA round trip number. (b) The noise values from Fig. 5(a) re-drawn as a function of the normalized gain and gain depletion. For high normalized gain and high gain depletion, a highly stable operation point can be observed. (c) To explain the mapping from Fig. 5(a) to Fig. 5(b), The noise for operation along the grey dashed lines in Fig. 5(a) is re-drawn as a function of the normalized gain and gain depletion. The trajectories (1), (2), and (3) represent operation in regimes ①, ②, and ③. (d) Gain dynamics for Ho:YLF RA operated in regime ④. Stable output pulses can only be extracted at low gain depletion values.

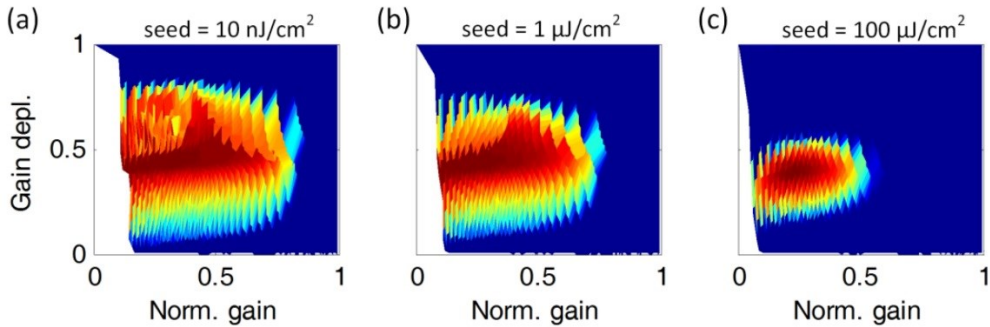

Fig. 6. Output noise as a function of the normalized gain and gain depletion for the seed fluences (a) 10 nJ/cm$^2$, (b) 1 μJ/cm$^2$, and (c) 100 μJ/cm$^2$. With the increase in seed energy, the region increases in which stable operation is possible. Stability increases from blue to red. Deep blue represent the absence of pulse instability.

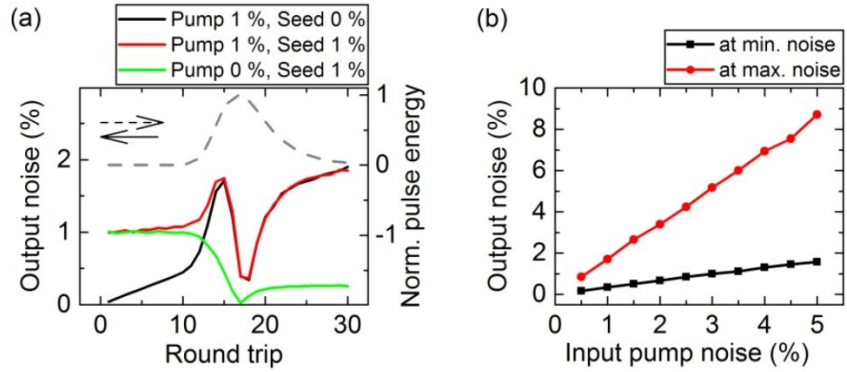

Fig. 7. (a) Output pulse noise and output pulse energy as a function of RT for three combinations of pump and seed noise. (b) Output pulse noise at RT=15 (maximum noise) and RT=18 (minimum noise) as a function of the input pump noise.

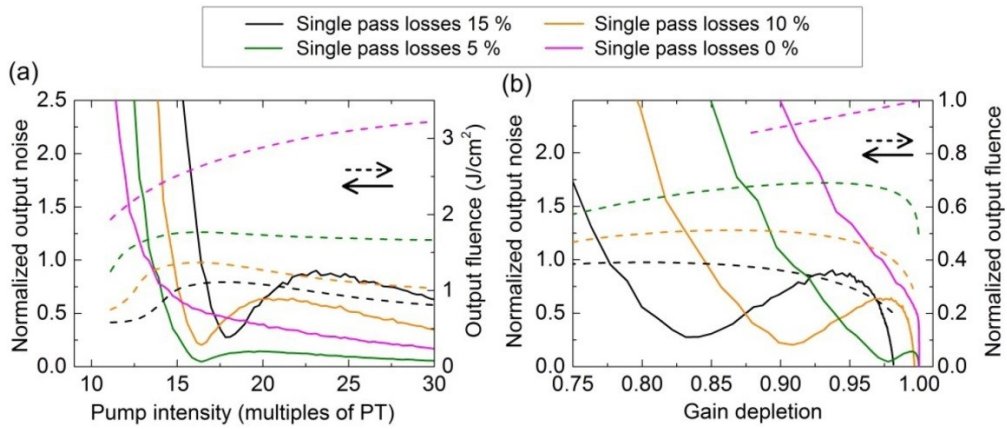

Fig. 8. (a) Normalized output noise as a function of the pump intensity. (b) The same graph re-drawn as a fuction of the gain depletion. The output fluence in (a) was additionally normalized in (b) to the stored fluence in the gain medium [Eq. (16)]. The simulations were conducted with different numbers of single-pass losses at a normalized repetition rate of 15. Only the RAs with losses show an intermediate noise minimum.

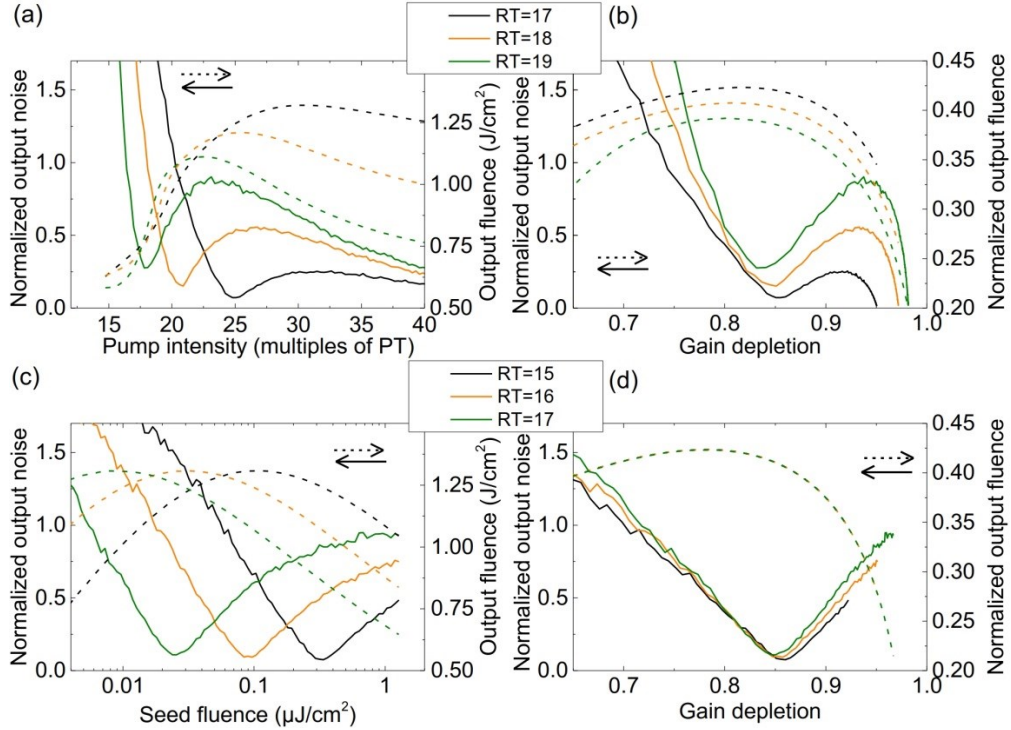

Fig. 9. Normalized output noise and pulse fluence as a function of (a) the pump intensity and (c) the seed fluence at a normalized repetition rate of 15. Re-plot of (a) and (c) as a function of the gain depletion in (b) and (d), respectively. The output fluence in (a) and (c) was additionally normalized in (b) and (d) to the stored fluence in the gain medium [Eq. (16)].

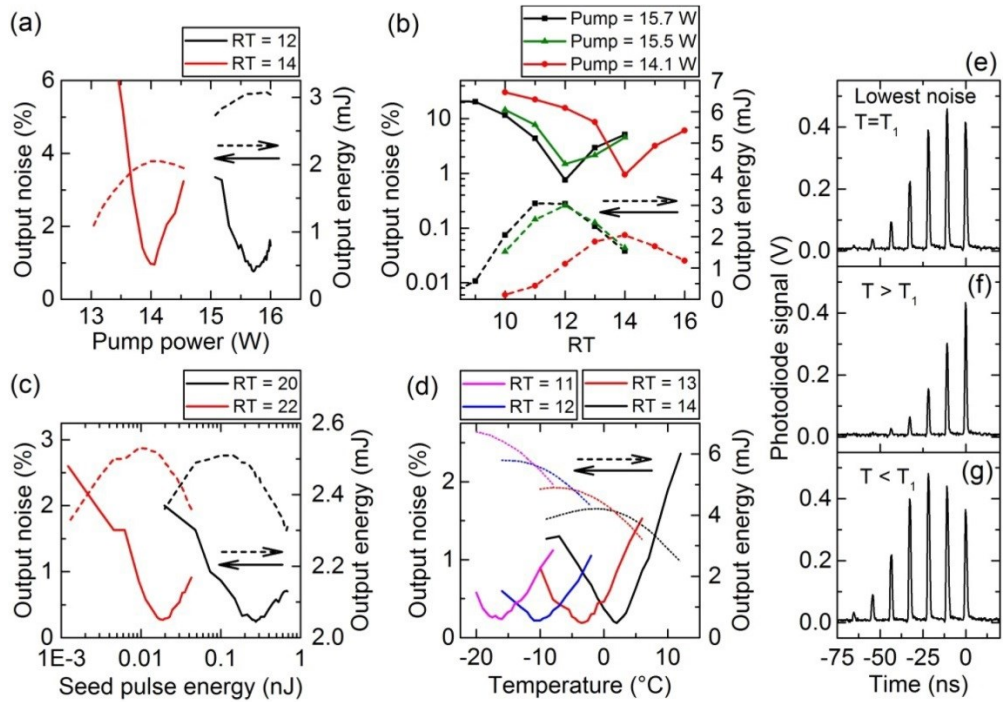

Fig. 10. Measured output noise and pulse energy as a function of (a) the pump power, (b) the RT, (c) the seed pulse energy, and (d) the crystal holder temperature. Intra-cavity pulse built-up for operation at, before and after the noise minimum in (e), (f) and (g), respectively, controlled by the crystal holder temperature.

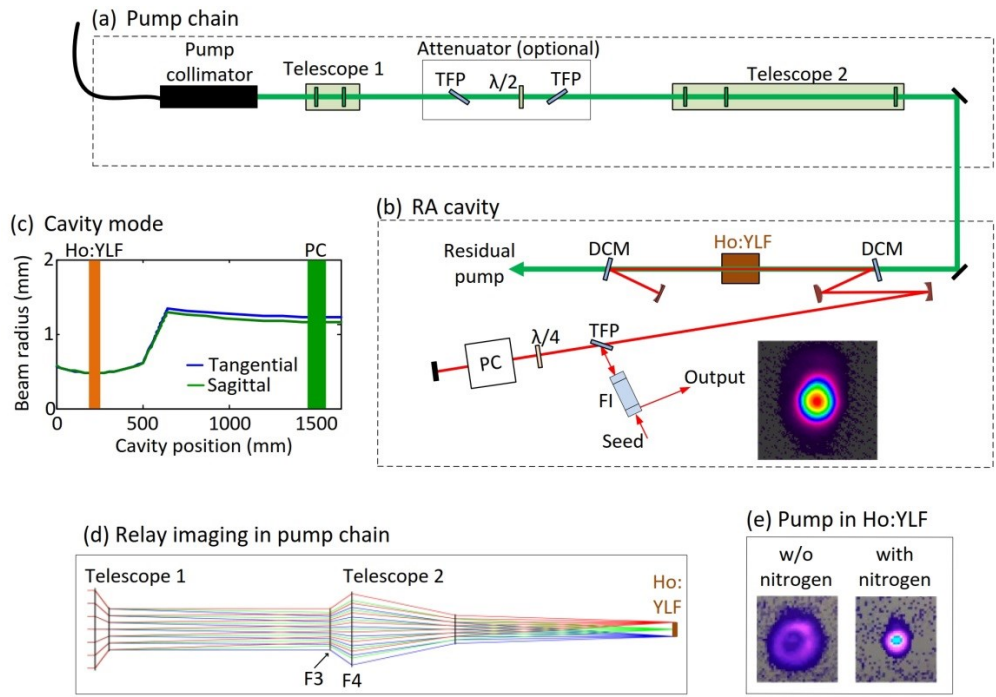

Fig. 11. Schematic of the Ho:YLF RA setup. (a) The pump chain consists of a polarization splitting attenuator (optional) and two telescopes that adjust the image size of the pump fiber in the Ho:YLF crystal. (b) The linear RA cavity. (b) The simulated cavity modes show good overlap between the sagittal and tangential plane. (d) Simulated path of rays of the pump laser, creating an image of the end of the pump fiber in the Ho:YLF crystal. (e) Typical pump beam spot in the Ho:YLF crystal plane for the RA setup under laboratory atmosphere and for the setup purged with nitrogen. DCM, dichroic mirror; TFP, thin film polarizer; λ/4-wave plate; PC, Pockels cell; FI, Faraday isolator.